\documentclass[twocolumn,british,aps,pra,showpacs,showkeys]{revtex4-1}
\usepackage{lmodern}
\usepackage[T1]{fontenc}
\usepackage[latin9]{inputenc}
\usepackage{babel}

\usepackage{amsthm}
\usepackage{amsmath}
\usepackage{graphicx}
\usepackage{amssymb}
\usepackage[unicode=true, 
 bookmarks=true,bookmarksnumbered=false,bookmarksopen=false,
 breaklinks=false,pdfborder={0 0 0},backref=false,colorlinks=false]
 {hyperref}
\hypersetup{pdftitle={Linking a distance measure of entanglement to its convex roof},
 pdfauthor={Alexander Streltsov, Hermann Kampermann, Dagmar Bruß},
 pdfkeywords={quantum information, entanglement measures}}
 
\makeatletter
\@ifundefined{textcolor}{}
{%
 \definecolor{BLACK}{gray}{0}
 \definecolor{WHITE}{gray}{1}
 \definecolor{RED}{rgb}{1,0,0}
 \definecolor{GREEN}{rgb}{0,1,0}
 \definecolor{BLUE}{rgb}{0,0,1}
 \definecolor{CYAN}{cmyk}{1,0,0,0}
 \definecolor{MAGENTA}{cmyk}{0,1,0,0}
 \definecolor{YELLOW}{cmyk}{0,0,1,0}
 }
\theoremstyle{plain}
\newtheorem{thm}{Theorem}
  \theoremstyle{plain}
  \newtheorem{prop}{Proposition}
  \theoremstyle{plain}
  \newtheorem{cor}{Corollary}

\usepackage{lmodern}
\usepackage{babel}
\usepackage{braket}

\makeatother

\begin{document}

\title{Linking a distance measure of entanglement to its convex roof}

\author{Alexander Streltsov}

\email{streltsov@thphy.uni-duesseldorf.de}

\author{Hermann Kampermann}

\email{kampermann@thphy.uni-duesseldorf.de}

\author{Dagmar Bruß}

\email{bruss@thphy.uni-duesseldorf.de}

\affiliation{Heinrich-Heine-Universität Düsseldorf, Institut für Theoretische
Physik III, D-40225 Düsseldorf, Germany}


\begin{abstract}
An important problem in quantum information theory is the quantification of entanglement in multipartite mixed quantum states. In this work, a connection between the geometric measure of entanglement and a distance measure of entanglement is established. We present a new expression for the geometric measure of entanglement in terms of the maximal fidelity with a separable state. A direct application of this result provides a closed expression for the Bures measure of entanglement of two qubits. We also prove that the number of elements in an optimal decomposition w.r.t. the geometric measure of entanglement is bounded from above by the Caratheodory bound, and we find necessary conditions for the structure of an optimal decomposition. 
\end{abstract}
\maketitle

\section{Introduction}

Entanglement \cite{Einstein1935} is one of the most fascinating features
of quantum mechanics, and allows a new view on information processing.
In spite of the central role of entanglement there does not yet exist
a complete theory for its quantification. Various entanglement measures
have been suggested - for an overview see \cite{Horodecki2009,Plenio2007}.

A composite pure quantum state $\ket{\psi}$ is called entangled iff
it can not be written as a product state. A composite mixed quantum
state $\rho$ on a Hilbert space $\mathcal{H}=\otimes_{j=1}^{n}\mathcal{H}_{j}$
is called entangled iff it cannot be written in the form \cite{Werner1989,Horodecki2009}
\begin{equation}
\rho=\sum_{i}p_{i}\left(\otimes_{j=1}^{n}\ket{\psi_{i}^{\left(j\right)}}\bra{\psi_{i}^{\left(j\right)}}\right)\label{eq:sep}\end{equation}
 with $p_{i}>0$, $\sum_{i}p_{i}=1$, and where $n\geq2$ and $\ket{\psi_{i}^{\left(j\right)}}\in\mathcal{H}_{j}$.

The degree of entanglement can be captured in a function $E\left(\rho\right)$
that should fulfil at least the following criteria \cite{Horodecki2009}: 
\begin{itemize}
\item $E\left(\rho\right)\geq0$ and equality holds iff $\rho$ is separable
\footnote{Note that the distillable entanglement $E_{D}$ does not satisfy this
criterion, i.e. it can be zero on entangled states. However it is
also accepted as a measure of entanglement \cite{Horodecki2009}.%
},
\item $E$ cannot increase under local operations and classical communication
(LOCC), i.e. $E\left(\Lambda\left(\rho\right)\right)\leq E\left(\rho\right)$
for any LOCC map $\Lambda$. 
\end{itemize}
These criteria are satisfied by all measures of entanglement presented
in this paper. One possibility to define an entanglement measure for
a mixed quantum state $\rho$ is via its {\em distance} to the
set of separable states \cite{Vedral1997}, for an illustration see
Figure \ref{fig:D}. Another possibility to define an entanglement
measure for a mixed quantum state $\rho$ is the {\em convex roof}
extension, in which the entanglement is quantified by the weighted
sum of the entanglement measure of the pure states in a given decomposition
of $\rho$, minimised over all possible decompositions. There is no
{\em a priori} reason why these two types of entanglement measures
should be related. In this paper we will establish a link between
them, by showing the equality between the convex roof extension of
the geometric measure of entanglement for pure states, and the corresponding
distance measure based on the fidelity with the closest separable
state. Using this result, we will also study the properties of the
optimal decompositions of the given state $\rho$, and its closest
separable state.

Our paper is organised as follows: In section \ref{sec:Definitions}
we provide the definitions of the used entanglement measures. In section
\ref{sec:GME} we derive a main result of this paper, namely the equality
between the convex roof extension of the geometric measure of entanglement
and the fidelity-based distance measure. In section \ref{sec:Two-qubits}
we study the most simple composite quantum system, namely two qubits,
give an analytical expression for the Bures measure of entanglement,
and consider other measures that are based on the geometric measure
of entanglement. In section \ref{sec:Optimal-decompositions} we characterise
the optimal decomposition of $\rho$ (i.e. the one that reaches the
minimum in the convex roof construction) from knowledge of the closest
separable state and vice versa. Finally, in section \ref{sec:Structure}
we derive a necessary criterion that the states in an optimal decomposition
have to fulfil. We conclude in section \ref{sec:Concluding-remarks}.

\section{\label{sec:Definitions}Definitions}

\begin{figure}
\includegraphics[width=0.5\columnwidth]{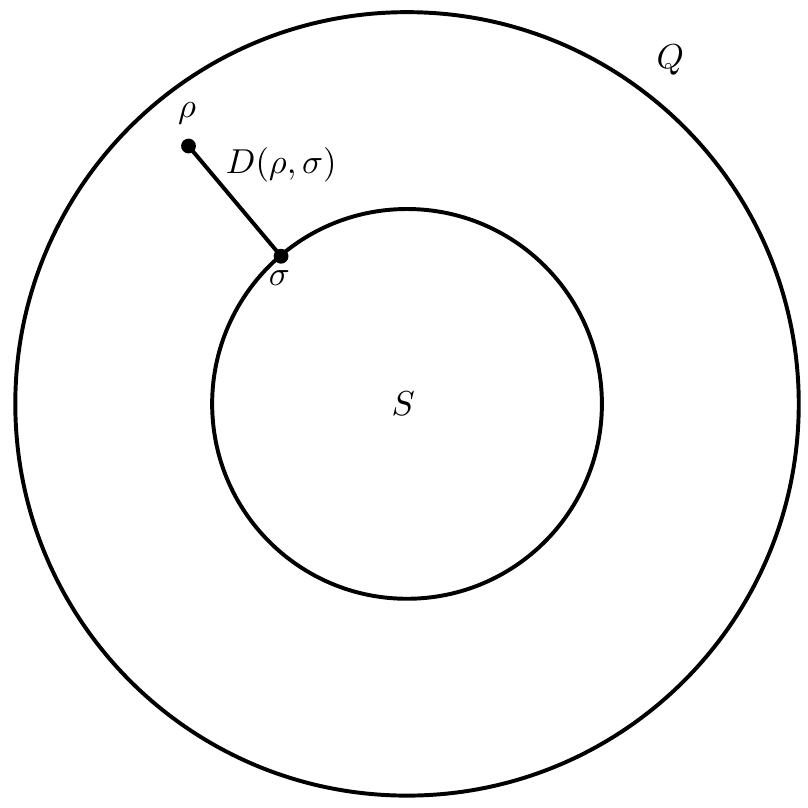}

\caption{\label{fig:D}$S$ denotes the set of separable states within the
set of all quantum states $Q$. The state $\sigma$ is the closest
separable state to $\rho$, with respect to the distance $D$. }

\end{figure}

Two classes of entanglement measures are considered in this paper.
The first class consists of measures based on a distance \cite{Vedral1997,Vedral1998}:
\begin{equation}
E_{D}\left(\rho\right)=\inf_{\sigma\in S}D\left(\rho,\sigma\right),\end{equation}
 where $D\left(\rho,\sigma\right)$ is a {}``distance'' between
$\rho$ and $\sigma$ and $S$ is the set of separable states. This
concept is illustrated in Figure \ref{fig:D}. Following \cite{Horodecki2009},
we do not require a distance to be a metric. In this paper we will
consider for example the Bures measure of entanglement \cite{Vedral1998}:
\begin{equation}
E_{B}\left(\rho\right)=\min_{\sigma\in S}\left(2-2\sqrt{F\left(\rho,\sigma\right)}\right),\label{eq:bures}\end{equation}
 where $F\left(\rho,\sigma\right)=\left(\mathrm{Tr}\left[\sqrt{\sqrt{\rho}\sigma\sqrt{\rho}}\right]\right)^{2}$
is Uhlmann's fidelity \cite{Uhlmann1976}. A very similar measure
is the Groverian measure of entanglement \cite{Biham2002,Shapira2006},
defined as \begin{equation}
E_{Gr}\left(\rho\right)=\min_{\sigma\in S}\sqrt{1-F\left(\rho,\sigma\right)}\ .\label{groverian}\end{equation}
 As it can be expressed as a simple function of $E_{B}$, we will
not consider it explicitly. Another important representant of the
first class is the relative entropy of entanglement defined as \cite{Vedral1998}
\begin{equation}
E_{R}\left(\rho\right)=\min_{\sigma\in S}S\left(\rho||\sigma\right),\end{equation}
 where $S\left(\rho||\sigma\right)$ is the relative entropy: \begin{equation}
S\left(\rho||\sigma\right)=\mathrm{Tr}\left[\rho\log_{2}\rho\right]-\mathrm{Tr}\left[\rho\log_{2}\sigma\right].\end{equation}

The second class of entanglement measures consists of convex roof
measures \cite{Uhlmann1997}: \begin{eqnarray}
E\left(\rho\right) & = & \min\sum_{i}p_{i}E\left(\ket{\psi_{i}}\right),\end{eqnarray}
 where $\sum_{i}p_{i}=1$, $p_{i}\geq0$, and the minimum is taken
over all pure state decompositions of $\rho=\sum_{i}p_{i}\ket{\psi_{i}}\bra{\psi_{i}}$.
An important example of the second class is the geometric measure
of entanglement $E_{G}$, defined as follows \cite{Wei2003}: \begin{eqnarray}
E_{G}\left(\ket{\psi}\right) & = & 1-\max_{\ket{\phi}\in S}\left|\braket{\phi|\psi}\right|^{2},\\
E_{G}\left(\rho\right) & = & \min\sum_{i}p_{i}E_{G}\left(\ket{\psi_{i}}\right),\label{geomconvex}\end{eqnarray}
where the minimum is taken over all pure state decompositions of $\rho$.
Entanglement measures of this form were considered earlier in \cite{Shimony1995}
and \cite{Barnum2001}. Another important representant of the second
class for bipartite states $\rho^{AB}$ is the entanglement of formation
$E_{F}$, which is for pure states $\rho=\ket{\psi}\bra{\psi}$ defined
as the von Neumann entropy of the reduced density matrix, \begin{equation}
E_{F}\left(\ket{\psi}\right)=-\mathrm{Tr}\left[\rho^{A}\log_{2}\rho^{A}\right],\end{equation}
 where $\rho^{A}=\mathrm{Tr}_{B}\left[\ket{\psi}\bra{\psi}\right]$.
For mixed states this measure is again defined via the convex roof
construction \cite{Bennett1996}: \begin{equation}
E_{F}\left(\rho\right)=\min_{\left\{ p_{i},\ket{\psi_{i}}\right\} }\sum_{i}p_{i}E_{F}\left(\ket{\psi_{i}}\right).\end{equation}
 For two-qubit states analytic formulae for $E_{F}$ and $E_{G}$
are known; both are simple functions of the Concurrence \cite{Wootters1998,Wei2003}.

Remember that the Concurrence for a two-qubit state $\rho$ is given
by \cite{Wootters1998} \begin{equation}
C(\rho)=\text{max}\{\xi_{1}-\xi_{2}-\xi_{3}-\xi_{4},0\}\ ,\end{equation}
 where $\xi_{i}$, with $i\in\{1,2,3,4\}$, are the square roots of
the eigenvalues of $\rho\cdot\tilde{\rho}$ in decreasing order, and
$\tilde{\rho}$ is defined as $\tilde{\rho}=(\sigma_{y}\otimes\sigma_{y})\rho^{*}(\sigma_{y}\otimes\sigma_{y})$.

The entanglement of formation for a two-qubit state $\rho$ as a function
of the concurrence is expressed as \cite{Wootters1998} \begin{equation}
E_{F}(\rho)=h(\frac{1}{2}+\frac{1}{2}\sqrt{1-C(\rho)^{2}})\ ,\end{equation}
 where $h(x)=-x\log_{2}x-(1-x)\log_{2}(1-x)$ is the Shannon entropy.
The geometric measure of entanglement for a two-qubit state $\rho$
as a function of the concurrence was shown in \cite{Wei2003} to be
\begin{equation}
E_{G}(\rho)=\frac{1}{2}(1-\sqrt{1-C(\rho)^{2}})\ .\label{eq:Eg-1}\end{equation}

This formula was already found in \cite{Vidal2000} in a different
context. For bipartite states it is furthermore known that \cite{Vedral1998}
\begin{equation}
E_{F}\left(\rho\right)\geq E_{R}\left(\rho\right),\end{equation}
 where for bipartite pure states the equal sign holds \cite{Vedral1998}.

The geometric measure of entanglement plays an important role in the
research of fundamental properties of quantum systems. Recently it
has been used to show that the most quantum states are too entangled
to be used for quantum computation \cite{Gross2009}. In \cite{Guhne2007}
the authors showed how a lower bound on the geometric measure of entanglement
can be estimated in experiments. A connection to Bell inequalities
for graph states has also been reported \cite{Guhne2005}.

\section{\label{sec:GME}Geometric measure of entanglement for mixed states}

In this section we will show a main result of our paper: the geometric
measure of entanglement, defined via the convex roof, see eq. (\ref{geomconvex}),
is equal to a distance-based alternative.

We introduce the \emph{fidelity of separability} \begin{equation}
F_{s}\left(\rho\right)=\max_{\sigma\in S}F\left(\rho,\sigma\right),\label{eq:fs}\end{equation}
 where the maximum is taken over all separable states of the form
\eqref{eq:sep}. 
\begin{thm}
\label{thm:main}For a multipartite mixed state $\rho$ on a finite
dimensional Hilbert space $\mathcal{H}=\otimes_{j=1}^{n}\mathcal{H}_{j}$
the following equality holds:\begin{equation}
F_{s}\left(\rho\right)=\max_{\left\{ p_{i},\ket{\psi_{i}}\right\} }\sum_{i}p_{i}F_{s}\left(\ket{\psi_{i}}\right),\end{equation}
 where the maximisation is done over all pure state decompositions
of $\rho=\sum_{i}p_{i}\ket{\psi_{i}}\bra{\psi_{i}}$.
\end{thm}
\begin{proof} Remember that according to Uhlmann's theorem \cite[page 411]{Nielsen2000}
\begin{equation}
F\left(\rho,\sigma\right)=\max_{\ket{\phi}}\left|\braket{\psi|\phi}\right|^{2},\end{equation}
holds for two arbitrary states $\rho$ and $\sigma$, where $\ket{\psi}$
is a purification of $\rho$ and the maximisation is done over all
purifications of $\sigma$, which are denoted by $\ket{\phi}$.

We start the proof with eq. (\ref{eq:fs}). In order to find $F_{s}\left(\rho\right)$
we have to maximise $\left|\braket{\psi|\phi}\right|^{2}$ over all
purifications $\ket{\phi}$ of all separable states $\sigma=\sum_{j}q_{j}\ket{\phi_{j}}\bra{\phi_{j}}$,
where all $\ket{\phi_{j}}$ are separable.

The purifications of $\rho$ and $\sigma$ can in general be written
as \begin{eqnarray}
\ket{\psi'} & = & \sum_{i}\sqrt{p'_{i}}\ket{\psi'_{i}}\otimes\ket{i},\\
\ket{\phi'} & = & \sum_{j}\sqrt{q_{j}}\ket{\phi_{j}}\otimes U^{\dagger}\ket{j},\end{eqnarray}
where $\left\{ p'_{i},\ket{\psi'_{i}}\right\} $ is a fixed decomposition
of $\rho$, $\braket{k|l}=\delta_{kl}$ and $U$ is a unitary on the
ancillary Hilbert space spanned by the states $\left\{ \ket{i}\right\} $.
To see that all purifications of a separable state $\sigma=\sum_{j}q_{j}\ket{\phi_{j}}\bra{\phi_{j}}$
are of the form given by $\ket{\phi'}$, we start with an arbitrary
purification $\ket{\phi''}=\sum_{k}\sqrt{r_{k}}\ket{\alpha_{k}}\otimes\ket{k}$,
such that $\sigma=\sum_{k}r_{k}\ket{\alpha_{k}}\bra{\alpha_{k}}$
and $\braket{k|l}=\delta_{k,l}$. Further holds: $\sqrt{r_{k}}\ket{\alpha_{k}}=\sum_{j}u_{kj}\sqrt{q_{j}}\ket{\phi_{j}}$,
with $u_{kj}$ being elements of a unitary matrix \cite{Hughston1993}.
Using the last relation we get $\ket{\phi''}=\sum_{j}\sqrt{q_{j}}\ket{\phi_{j}}\otimes\ket{j'}$
with $\ket{j'}=\sum_{k}u_{kj}\ket{k}$. Thus we brought an arbitrary
purification of $\sigma$ to the form given by $\ket{\phi'}$.

In order to find $F_{s}\left(\rho\right)$ in the above parametrisation
we have to maximise the overlap $\left|\braket{\psi'|\phi'}\right|^{2}$
over all unitaries $U$, all probability distributions $\left\{ q_{i}\right\} $
and all sets of separable states $\left\{ \ket{\phi_{i}}\right\} $. 

We will now show, that we can also achieve $F_{s}\left(\rho\right)$
by maximising the overlap $\left|\braket{\psi|\phi}\right|^{2}$ of
the purifications \begin{eqnarray}
\ket{\psi} & = & \sum_{i}\sqrt{p_{i}}\ket{\psi_{i}}\otimes\ket{i},\\
\ket{\phi} & = & \sum_{j}\sqrt{q_{j}}\ket{\phi_{j}}\otimes\ket{j},\end{eqnarray}
where now the maximisation has to be done over all decompositions
$\left\{ p_{i},\ket{\psi_{i}}\right\} $ of the given state $\rho$,
all probability distributions $\left\{ q_{i}\right\} $ and all sets
of separable states $\left\{ \ket{\phi_{i}}\right\} $. To see how
this works we write the matrix $U$ in its elements, $U=\sum_{k,l}u_{kl}\ket{k}\bra{l}$,
and apply it in the overlap $\left|\braket{\psi'|\phi'}\right|^{2}$,
thus noting that the action of the unitary is equivalent to a transformation
of the set of unnormalised states $\left\{ \sqrt{p'_{i}}\ket{\psi'_{i}}\right\} $
to the new set $\left\{ \sqrt{p{}_{i}}\ket{\psi{}_{i}}\right\} $.
The connection between the two sets is given by the unitary: $\sqrt{p_{i}}\ket{\psi{}_{i}}=\sum_{j}u_{ij}\sqrt{p'_{j}}\ket{\psi'_{j}}$,
which is a transformation between two decompositions of the state
$\rho$, see also \cite[p.103f]{Nielsen2000}. The advantage of this
parametrisation is that now both purifications have the same orthogonal
states on the ancillary Hilbert space. 

We now do the maximisation of the overlap \begin{equation}
\left|\braket{\psi|\phi}\right|=\left|\sum_{i}\sqrt{q_{i}}\sqrt{p_{i}}\braket{\psi_{i}|\phi_{i}}\right|,\end{equation}
starting with the separable states $\left\{ \ket{\phi_{i}}\right\} $.
The optimal states can be chosen such that all terms $\braket{\psi_{i}|\phi_{i}}$
are real, positive and equal to $\sqrt{F_{s}\left(\ket{\psi_{i}}\right)}=\max_{\ket{\phi}\in S}\left|\braket{\psi_{i}|\phi}\right|$,
it is obvious that this choice is optimal. We also used the fact that
for pure states $\ket{\psi}$ it is enough to maximise over pure separable
states: $F_{s}\left(\ket{\psi}\right)=\max_{\ket{\phi}\in S}\left|\braket{\psi|\phi}\right|^{2}$.
To see this note that $F\left(\ket{\psi}\bra{\psi},\sigma\right)=\braket{\psi|\sigma|\psi}$.
Suppose now, the closest separable state to $\ket{\psi}$ is the mixed
state $\sigma$ with the separable decomposition $\sigma=\sum_{j}q_{j}\ket{\phi_{j}}\bra{\phi_{j}}$,
all $\ket{\phi_{j}}$ being separable. Without loss of generality
let $\left|\braket{\psi|\phi_{1}}\right|\geq\left|\braket{\psi|\phi_{j}}\right|$
be true for all $j$. Then holds: $F\left(\ket{\psi}\bra{\psi},\sigma\right)=\braket{\psi|\sigma|\psi}=\sum_{j}q_{j}\left|\braket{\psi|\phi_{j}}\right|^{2}\leq\sum_{j}q_{j}\left|\braket{\psi|\phi_{1}}\right|^{2}=\left|\braket{\psi|\phi_{1}}\right|^{2}$,
and thus $\ket{\phi_{1}}$ is a closest separable state to $\ket{\psi}$.

The maximisation over $\left\{ \ket{\phi_{i}}\right\} $ gives us
\begin{equation}
\max_{\left\{ \ket{\phi_{j}}\right\} }\left|\braket{\psi|\phi}\right|=\sum_{i}\sqrt{q_{i}}\sqrt{p_{i}}\sqrt{F_{s}\left(\ket{\psi_{i}}\right)}.\label{vectors}\end{equation}
 Now we do the optimisation over $q_{i}$. Using Lagrange multipliers
we get \begin{equation}
\sqrt{q_{i}}=\frac{\sqrt{p_{i}}\sqrt{F_{s}\left(\ket{\psi_{i}}\right)}}{\sqrt{\sum_{k}p_{k}F_{s}\left(\ket{\psi_{k}}\right)}},\end{equation}
 with the result \begin{equation}
\max_{\left\{ q_{j},\ket{\phi_{j}}\right\} }\left|\braket{\psi|\phi}\right|^{2}=\sum_{i}p_{i}F_{s}\left(\ket{\psi_{i}}\right).\end{equation}
 It is easy to understand that this choice of $\{q_{i}\}$ is optimal,
when one interprets the right hand side of eq. (\ref{vectors}) as
a scalar product between a vector with entries $(\sqrt{p_{1}}\sqrt{F_{s}\left(\ket{\psi_{1}}\right)},\sqrt{p_{2}}\sqrt{F_{s}\left(\ket{\psi_{2}}\right)},...)$
and a vector with entries $(\sqrt{q_{1}},\sqrt{q_{2}},...)$. The
scalar product of two vectors with given length is maximal when they
are parallel.

In the last step we do the maximisation over all decompositions $\left\{ p_{i},\ket{\psi_{i}}\right\} $
of the given state $\rho$ which leads to the end of the proof, namely
\begin{equation}
F_{s}\left(\rho\right)=\max\left|\braket{\psi|\phi}\right|^{2}=\max_{\left\{ p_{i},\ket{\psi_{i}}\right\} }\sum_{i}p_{i}F_{s}\left(\ket{\psi_{i}}\right).\label{eq:Fs}\end{equation}

\end{proof} We can generalise Theorem \ref{thm:main} for arbitrary
convex sets; the result can be found in Appendix \ref{sec:Convex-sets}.
Using Theorem \ref{thm:main} it follows immediately that the geometric
measure of entanglement is not only a convex roof measure, but also
a distance based measure of entanglement: 
\begin{prop}
\label{pro:Eg-F}For a multipartite mixed state $\rho$ on a finite
dimensional Hilbert space $\mathcal{H}=\otimes_{j=1}^{n}\mathcal{H}_{j}$
the following equality holds:\begin{equation}
E_{G}\left(\rho\right)=1-\max_{\sigma\in S}F\left(\rho,\sigma\right).\label{eq:Eg}\end{equation}

\end{prop}
Proposition \ref{pro:Eg-F} establishes a connection between $E_{G}$
and distance based measures like the Bures measure $E_{B}$ and Groverian
measure $E_{Gr}$. All of them are simple functions of each other.

In \cite{Wei2004} the authors found the following connection between
$E_{R}$ and $E_{G}$ for pure states:\begin{equation}
E_{R}\left(\ket{\psi}\right)\geq-\log_{2}\left(1-E_{G}\left(\ket{\psi}\right)\right).\label{eq:ER-1}\end{equation}
 This inequality can be generalised to mixed states as follows:\begin{equation}
E_{R}\left(\rho\right)\geq\max\left\{ 0,-\log_{2}\left(1-E_{G}\left(\rho\right)\right)-S\left(\rho\right)\right\} ,\label{eq:ER}\end{equation}
 where $S\left(\rho\right)=-\mathrm{Tr}\left[\rho\log_{2}\rho\right]$
is the von Neumann entropy of the state. The inequality \eqref{eq:ER}
is a direct consequence of the following proposition. %
\begin{figure}
\includegraphics[width=1\columnwidth]{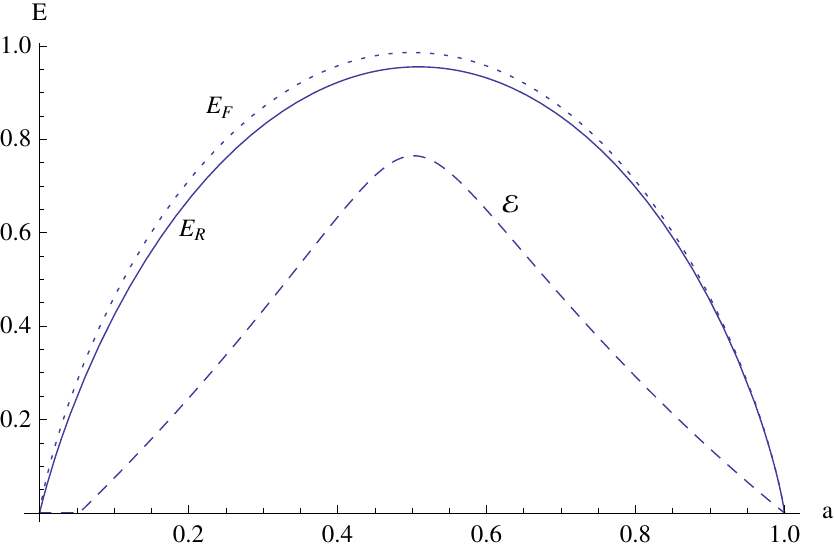}

\caption{\label{fig:GVP-1}Entanglement of formation $E_{F}$ (dotted curve),
relative entropy of entanglement $E_{R}$ (solid curve) and $\mathcal{E}=\max\left\{ 0,-\log_{2}\left(1-E_{G}\left(\rho\right)\right)-S\left(\rho\right)\right\} $
(dashed curve) of the state $\rho=p\ket{\psi}\bra{\psi}+\left(1-p\right)\ket{01}\bra{01}$
with $\ket{\psi}=\sqrt{a}\ket{01}+\sqrt{1-a}\ket{10}$ for $p=\frac{99}{100}$
as a function of $a$.}

\end{figure}

\begin{prop}
\label{pro:Re-F}For two arbitrary quantum states $\rho$ and $\sigma$
holds:\begin{eqnarray}
S\left(\rho||\sigma\right) & \geq & \mathrm{Tr}\left[\rho\log_{2}\rho\right]-\log_{2}F\left(\rho,\sigma\right).\label{eq:S}\end{eqnarray}

\end{prop}
\begin{proof} With $\rho=\sum_{i}p_{i}\ket{\psi_{i}}\bra{\psi_{i}}$
we will estimate $-\mathrm{Tr}\left[\rho\log_{2}\sigma\right]$ from
below: \begin{eqnarray}
-\mathrm{Tr}\left[\rho\log_{2}\sigma\right] & = & -\sum_{i}p_{i}\braket{\psi_{i}|\log_{2}\sigma|\psi_{i}}\\
 & \geq & -\sum_{i}p_{i}\log_{2}\braket{\psi_{i}|\sigma|\psi_{i}}.\end{eqnarray}
 Here we used concavity of the $\log$ function: \begin{equation}
\log_{2}\braket{\psi_{i}|\sigma|\psi_{i}}\geq\braket{\psi_{i}|\log_{2}\sigma|\psi_{i}}.\end{equation}
 Using concavity again we get $\sum_{i}p_{i}\log_{2}\braket{\psi_{i}|\sigma|\psi_{i}}\leq\log_{2}\sum_{i}p_{i}\braket{\psi_{i}|\sigma|\psi_{i}}$
and thus \begin{eqnarray}
-\mathrm{Tr}\left[\rho\log_{2}\sigma\right] & \geq & -\log_{2}\sum_{i}p_{i}\braket{\psi_{i}|\sigma|\psi_{i}}\\
 & = & -\log_{2}\mathrm{Tr}\left[\rho\sigma\right].\end{eqnarray}
 The fidelity can be bounded from below as follows: \begin{eqnarray}
F\left(\rho,\sigma\right) & = & \left(\mathrm{Tr}\left[\sqrt{\sqrt{\rho}\sigma\sqrt{\rho}}\right]\right)^{2}=\left(\sum_{i}\lambda_{i}\right)^{2}\\
 & \geq & \sum_{i}\lambda_{i}^{2}=\mathrm{Tr}\left[\sqrt{\rho}\sigma\sqrt{\rho}\right]=\mathrm{Tr}\left[\rho\sigma\right],\end{eqnarray}
 where $\lambda_{i}$ are the eigenvalues of the positive operator
$\sqrt{\sqrt{\rho}\sigma\sqrt{\rho}}$. \end{proof} 

\begin{figure}
\includegraphics[width=1\columnwidth]{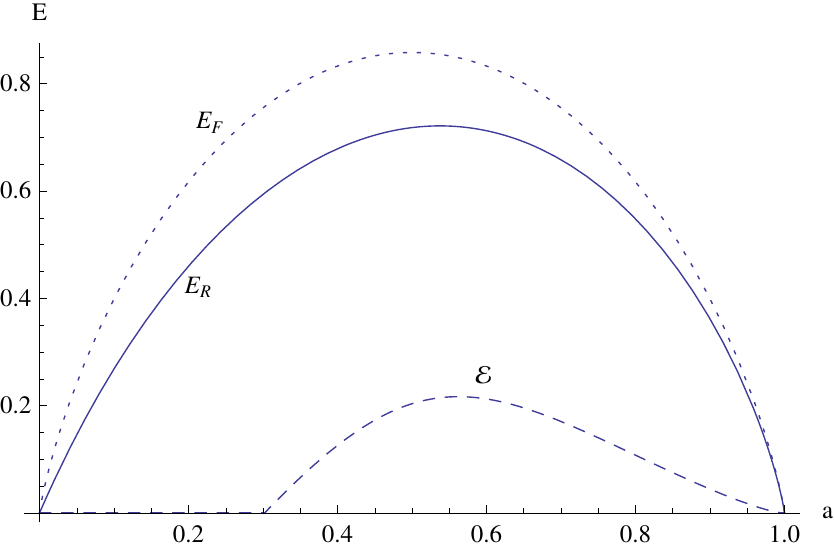}

\caption{\label{fig:GVP-2}Entanglement of formation $E_{F}$ (dotted curve),
relative entropy of entanglement $E_{R}$ (solid curve) and $\mathcal{E}=\max\left\{ 0,-\log_{2}\left(1-E_{G}\left(\rho\right)\right)-S\left(\rho\right)\right\} $
(dashed curve) of the state $\rho=p\ket{\psi}\bra{\psi}+\left(1-p\right)\ket{01}\bra{01}$
with $\ket{\psi}=\sqrt{a}\ket{01}+\sqrt{1-a}\ket{10}$ for $p=\frac{9}{10}$
as a function of $a$.}

\end{figure}
 The inequality (\ref{eq:ER}) becomes trivial for states with high
entropy. As a nontrivial example we consider the two qubit state \begin{equation}
\rho=p\ket{\psi}\bra{\psi}+\left(1-p\right)\ket{01}\bra{01},\label{eq:GVP}\end{equation}
with $\ket{\psi}=\sqrt{a}\ket{01}+\sqrt{1-a}\ket{10}$. This state
was called generalised Vedral-Plenio state in \cite{Miranowicz2008},
where the authors showed that the closest separable state $\sigma$
with respect to the relative entropy of entanglement is given by \begin{equation}
\sigma=\left(1-p+pa\right)\ket{01}\bra{01}+p\left(1-a\right)\ket{10}\bra{10}.\end{equation}
In Figure \ref{fig:GVP-1} and \ref{fig:GVP-2} we show the plot of
$E_{F}$ (dotted curve), $E_{R}$ (solid curve) and $\mathcal{E}=\max\left\{ 0,-\log_{2}\left(1-E_{G}\left(\rho\right)\right)-S\left(\rho\right)\right\} $
(dashed curve) as a function of $a$ for $p=\frac{99}{100}$ and $p=\frac{9}{10}$
respectively. It can be seen that $ $ $\mathcal{E}$ drops quickly
with increasing entropy of the state, and thus is nontrivial only
for states close to pure states with high entanglement.

In \cite{Plenio2000,Plenio2001} the authors gave lower bounds for
the relative entropy of entanglement in terms of the von Neumann entropies
of the reduced states, which provide better lower bounds for $E_{R}$
than \eqref{eq:ER}. Thus, the inequality \eqref{eq:ER} should be
seen as a connection between the two entanglement measures $E_{R}$
and $E_{G}$, and not as an improved lower bound for $E_{R}$.

\section{\label{sec:Two-qubits}Entanglement measures for two qubits}

\subsection{Bures measure of entanglement}

\begin{figure}
\includegraphics[width=1\columnwidth]{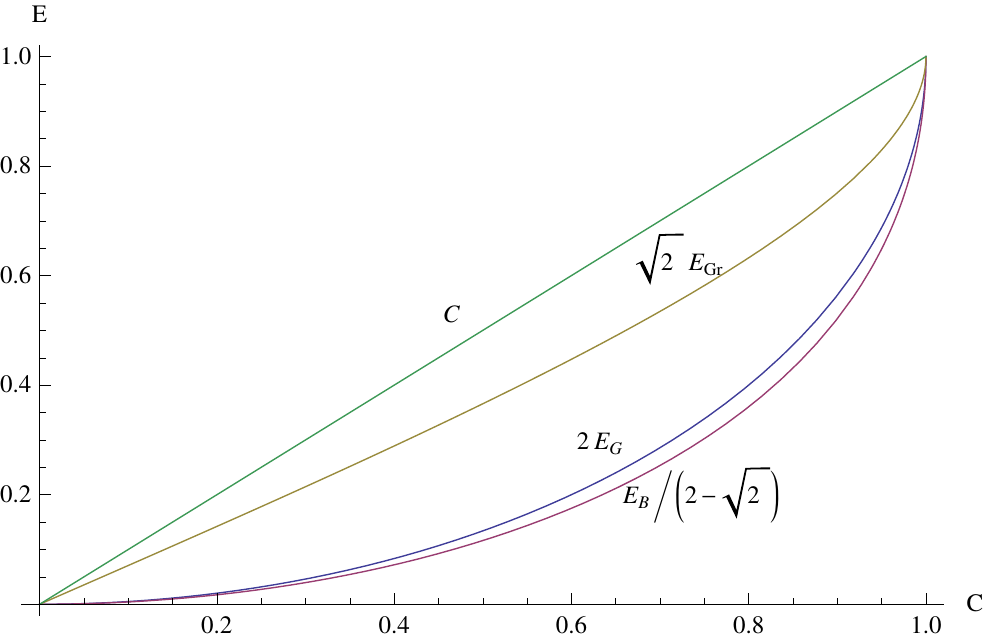} \caption{\label{fig:2} Plot of the geometric measure of entanglement $E_{G}$,
Bures measure of entanglement $E_{B}$ and Groverian measure of entanglement
$E_{Gr}$ as a function of the concurrence $C$ for two qubit states.
All measures were renormalised such that they reach $1$ for maximally
entangled states.}

\end{figure}

We can use Proposition \ref{pro:Eg-F} to evaluate entanglement measures
for two qubit states. From \cite{Vidal2000,Wei2003} we know the geometric
measure for two-qubit states as a function of the concurrence, see
eq. \eqref{eq:Eg-1}. Using this together with eq. \eqref{eq:Eg}
we find the fidelity of separability as function of the concurrence:
\begin{equation}
F_{s}(\rho)=\max_{\sigma\in S}F\left(\rho,\sigma\right)=\frac{1}{2}\left(1+\sqrt{1-C\left(\rho\right)^{2}}\right).\label{eq:Fs-2}\end{equation}
 Now we are able to give an expression for the Bures measure of entanglement
for two qubit states, remember its definition in eq. \eqref{eq:bures}.
\begin{prop} \label{pro:Bures}For any two qubit state $\rho$ the
Bures measure of entanglement is given by \begin{equation}
E_{B}\left(\rho\right)=2-2\sqrt{\frac{1+\sqrt{1-C\left(\rho\right)^{2}}}{2}}.\end{equation}
 \end{prop} Note that for a maximally entangled state $E_{G}=\frac{1}{2}$
and $E_{B}=2-\sqrt{2}$. In order to compare these measures we renormalise
them such that each of them becomes equal to $1$ for maximally entangled
states. We show the result in Figure \ref{fig:2}. There we also plot
the Groverian measure of entanglement, see eq. (\ref{groverian}).

\subsection{Measures induced by the geometric measure of entanglement}

We consider now any generalised measure of entanglement for two qubit
states $\rho$ which can be written as a function of the geometric
measure of entanglement: \begin{equation}
E_{f}\left(\rho\right)=f\left(E_{G}\left(\rho\right)\right).\end{equation}

\begin{prop} \label{pro:generalized} Let $f\left(x\right)$ be any
convex function that is nonnegative for $x\geq0$ and obeys $f\left(0\right)=0$.
Then for two qubits $E_{f}(\rho)=f\left(E_{G}\left(\rho\right)\right)$
is equal to its convex roof, that is \begin{equation}
E_{f}\left(\rho\right)=\min\sum_{i}p_{i}E_{f}\left(\ket{\psi_{i}}\right)=f\left(\frac{1}{2}\left(1-\sqrt{1-C\left(\rho\right)^{2}}\right)\right),\label{eq:Ef}\end{equation}
 where the minimisation is done over all pure state decompositions
of $\rho$.\end{prop} \begin{proof} From \cite{Wei2003} we know
that the geometric measure of entanglement is a convex nonnegative
function of the concurrence, see also \eqref{eq:Eg-1} and Figure
\ref{fig:2}. As shown in \cite{Wei2003}, from convexity follows
that $E_{G}$ and $E_{F}$ have identical optimal decompositions,
and every state in this optimal decomposition has the same concurrence.
This observation led directly to the expression \eqref{eq:Eg-1} for
$E_{G}$ of two qubit states.

As $f$ is convex, $E_{f}$ also is a convex function of the concurrence.
To see this we note that convexity of $E_{G}$ implies \begin{equation}
E_{G}\left(\sum_{i}p_{i}C_{i}\right)\leq\sum_{i}p_{i}E_{G}\left(C_{i}\right),\end{equation}
 where we defined $E_{G}\left(C\right)=\frac{1}{2}(1-\sqrt{1-C^{2}})$.
As $f\left(x\right)$ is convex, nonnegative and $f\left(0\right)=0$,
it also must be monotonously increasing for $x\geq0$. Thus we have
\begin{equation}
f\left(E_{G}\left(\sum_{i}p_{i}C_{i}\right)\right)\leq f\left(\sum_{i}p_{i}E_{G}\left(C_{i}\right)\right).\end{equation}
 Now we can use convexity of $f$ to get \begin{equation}
f\left(E_{G}\left(\sum_{i}p_{i}C_{i}\right)\right)\leq\sum_{i}p_{i}f\left(E_{G}\left(C_{i}\right)\right).\end{equation}
 Defining $E_{f}\left(C\right)=f\left(E_{G}\left(C\right)\right)=f\left(\frac{1}{2}\left(1-\sqrt{1-C^{2}}\right)\right)$
the inequality above becomes \begin{equation}
E_{f}\left(\sum_{i}p_{i}C_{i}\right)\leq\sum_{i}p_{i}E_{f}\left(C_{i}\right).\end{equation}
 This proves that $E_{f}\left(C\right)$ is a convex function of the
concurrence. Using the same argumentation as was used in \cite{Wei2003}
to prove the expression \eqref{eq:Eg-1} we see that \eqref{eq:Ef}
must hold. \end{proof} As an example consider the Bures measure of
entanglement which can be written as $E_{B}\left(\rho\right)=E_{f}\left(\rho\right)$
with the convex function $f=2-2\sqrt{1-E_{G}\left(\rho\right)}$.
Using Proposition \ref{pro:generalized} we see that for two qubits
the Bures measure of entanglement is equal to its convex roof.

However, this might not be the case for a general higher-dimensional
state $\rho$. To see this assume that $E_{B}\left(\rho\right)$ is
equal to $\min\sum_{i}p_{i}E_{B}\left(\ket{\psi_{i}}\right)$. This
means that $\sqrt{F_{s}\left(\rho\right)}$ is equal to $\max\sum_{i}p_{i}\sqrt{F_{s}\left(\ket{\psi_{i}}\right)}$.
On the other hand, from Theorem \ref{thm:main} we know that \begin{equation}
F_{s}\left(\rho\right)=\max\sum_{i}p_{i}F_{s}\left(\ket{\psi_{i}}\right),\end{equation}
 and using monotonicity and concavity of the square root we see: \begin{equation}
\sqrt{F_{s}\left(\rho\right)}=\max\sqrt{\sum_{i}p_{i}F_{s}\left(\ket{\psi_{i}}\right)}\ge\max\sum_{i}p_{i}\sqrt{F_{s}\left(\ket{\psi_{i}}\right)}.\label{eq:generalized}\end{equation}
 The Bures measure of entanglement is equal to its convex roof if
and only if the inequality \eqref{eq:generalized} becomes an equality
for all states $\rho$.

Finally we note, that any entanglement measure $E_{h}$ defined as
$E_{h}\left(\rho\right)=\min_{\sigma\in S}h\left(F\left(\rho,\sigma\right)\right)$
with a monotonously decreasing nonnegative function $h$, $h\left(1\right)=0$,
becomes $E_{h}\left(\rho\right)=h\left(F_{s}\left(\rho\right)\right)$,
and can be evaluated exactly for two qubits using Proposition \ref{pro:Eg-F}.
An example of such a measure is the Bures measure of entanglement.

\section{\label{sec:Optimal-decompositions}Optimal decompositions w.r.t.
geometric measure of entanglement and consequences for closest separable
states}

Let $\rho$ be an $n$-partite quantum state acting on a finite-dimensional
Hilbert space $\mathcal{H}=\otimes_{i=1}^{n}\mathcal{H}_{i}$ of dimension
$d$. A decomposition of a mixed state $\rho$ is a set $\left\{ p_{i},\ket{\psi_{i}}\right\} $
with $p_{i}>0$, $\sum_{i}p_{i}=1$, and $\rho=\sum_{i}p_{i}\ket{\psi_{i}}\bra{\psi_{i}}$.
Throughout this paper we will call a decomposition optimal if it minimises
the geometric measure of entanglement, i.e. if $E_{G}\left(\rho\right)=\sum_{i}p_{i}E_{G}\left(\ket{\psi_{i}}\right)$.
A separable state $\sigma$ is a closest separable state to $\rho$
if $E_{G}\left(\rho\right)=1-F\left(\rho,\sigma\right)$. In the following
we will show how to find an optimal decomposition of $\rho$, given
a closest separable state.

\subsection{Equivalence between closest separable states and optimal decompositions}

In the maximisation of $F\left(\rho,\sigma\right)$ we can restrict
ourselves to separable states $\sigma$ acting on the same Hilbert
space $\mathcal{H}$. To see this, note that this is obviously true
for pure states, as we can always find a pure separable state $\ket{\phi}\in\mathcal{H}$
such that $\left|\braket{\psi|\phi}\right|^{2}$ is maximal. (Extra
dimensions cannot increase the overlap with the original state.) Let
now $\sigma=\sum_{i}q_{i}\ket{\phi_{i}}\bra{\phi_{i}}$ be a closest
separable state with purification $\ket{\phi}$ such that $F_{s}\left(\rho\right)=\left|\braket{\psi|\phi}\right|^{2}$,
where $\ket{\psi}$ is a purification of $\rho$. We can again write
the purifications as \begin{eqnarray}
\ket{\psi} & = & \sum_{i}\sqrt{p_{i}}\ket{\psi_{i}}\ket{i},\\
\ket{\phi} & = & \sum_{j}\sqrt{q_{j}}\ket{\phi_{j}}\ket{j},\end{eqnarray}
 with separable pure states $\ket{\phi_{j}}$ such that $\sqrt{F_{s}\left(\ket{\psi_{i}}\right)}=\braket{\psi_{i}|\phi_{i}}$.
As the states $\ket{\phi_{j}}$ are elements of $\mathcal{H}$, the
reduced state $\sigma=\mathrm{Tr}_{a}\left[\ket{\phi}\bra{\phi}\right]$
is a bounded operator acting on the same Hilbert space $\mathcal{H}$,
$\mathrm{Tr}_{a}$ denotes partial trace over the ancillary Hilbert
space spanned by the orthonormal basis $\left\{ \ket{i}\right\} $.

Now we are in position to prove the following result. \begin{prop}
\label{pro:state-decomposition} Let $\rho$ be an $n$-partite quantum
state acting on $\mathcal{H}=\otimes_{i=1}^{n}\mathcal{H}_{i}$. The
separable state $\sigma=\sum_{j=1}^{s}q_{j}\ket{\phi_{j}}\bra{\phi_{j}}$
with $s\geq d$ separable pure states $\ket{\phi_{j}}$ and $\sum_{j=1}^{s}q_{j}=1$,
$q_{i}\geq0$, is the closest separable state if and only if there
exists an optimal decomposition $\left\{ p_{i},\ket{\psi_{i}}\right\} _{i=1}^{s}$
with $s\geq d$ elements such that holds: $\sqrt{F_{s}\left(\ket{\psi_{i}}\right)}=\braket{\psi_{i}|\phi_{i}}$
and $q_{i}=\frac{p_{i}F_{s}\left(\ket{\psi_{i}}\right)}{\sum_{k}p_{k}F_{s}\left(\ket{\psi_{k}}\right)}$.
\end{prop} \begin{proof} In the following $\left\{ \ket{i}\right\} $
denotes a basis on the ancillary Hilbert space $\mathcal{H}_{a}$.
The closest separable state $\sigma=\sum_{j=1}^{s}q_{j}\ket{\phi_{j}}\bra{\phi_{j}}$
can be purified by \begin{equation}
\ket{\phi}=\sum_{j=1}^{s}\sqrt{q_{j}}\ket{\phi_{j}}\ket{j}.\label{eq:phi}\end{equation}
 We write a purification of the state $\rho$ as \begin{equation}
\ket{\psi}=\sum_{i=1}^{s}\sqrt{\lambda_{i}}\ket{\lambda_{i}}U\ket{i},\label{eq:psi}\end{equation}
 where $\lambda_{i}$ are the eigenvalues and $\ket{\lambda_{i}}$
are the corresponding eigenstates of $\rho$, with $\lambda_{i}=0$
for $i\geq d$, and $U$ is a unitary acting on the ancillary Hilbert
space $\mathcal{H}_{a}$. According to Uhlmann's theorem \cite{Uhlmann1976,Nielsen2000}
it holds: \begin{equation}
\left|\braket{\psi|\phi}\right|^{2}\leq F\left(\rho,\sigma\right)=F_{s}\left(\rho\right).\label{eq:Uhlmann}\end{equation}
 In the following let $U$ be a unitary such that equality is achieved
in \eqref{eq:Uhlmann}; its existence is assured by Uhlmann's theorem.
Writing $U=\sum_{k,l=1}^{s}u_{kl}\ket{k}\bra{l}$ in \eqref{eq:psi}
we get:\begin{eqnarray}
\ket{\psi} & = & \sum_{k,l=1}^{s}u_{kl}\sqrt{\lambda_{l}}\ket{\lambda_{l}}\ket{k}=\sum_{k=1}^{s}\sqrt{p_{k}}\ket{\psi_{k}}\ket{k}\label{eq:psi-1}\end{eqnarray}
 with $\sqrt{p_{k}}\ket{\psi_{k}}=\sum_{l=1}^{s}u_{kl}\sqrt{\lambda_{l}}\ket{\lambda_{l}}$.
Note that $\left\{ p_{k},\ket{\psi_{k}}\right\} _{k=1}^{s}$ is a
decomposition of $\rho$.

We will now show that $\left\{ p_{k},\ket{\psi_{k}}\right\} _{k=1}^{s}$
is an optimal decomposition by showing that $\left|\braket{\psi|\phi}\right|^{2}=\sum_{i}p_{i}F_{s}\left(\ket{\psi_{i}}\right)$.
As we chose the purifications such that $\left|\braket{\psi|\phi}\right|^{2}=F_{s}\left(\rho\right)$,
this will complete the proof. Computing the overlap $\left|\braket{\psi|\phi}\right|^{2}$
using \eqref{eq:phi} and \eqref{eq:psi-1} we get: \begin{equation}
\left|\braket{\psi|\phi}\right|^{2}=\left|\sum_{i}\sqrt{p_{i}q_{i}}\braket{\psi_{i}|\phi_{i}}\right|^{2}.\label{eq:overlap}\end{equation}
 As in the proof of Theorem \ref{thm:main}, maximality of \eqref{eq:overlap}
implies that $\left|\braket{\psi_{i}|\phi_{i}}\right|=\sqrt{F_{s}\left(\ket{\psi_{i}}\right)}$
and $q_{i}=\frac{p_{i}F_{s}\left(\ket{\psi_{i}}\right)}{\sum_{k}p_{k}F_{s}\left(\ket{\psi_{k}}\right)}$.
Then we immediately see that $\left\{ p_{k},\ket{\psi_{k}}\right\} _{k=1}^{s}$
is optimal, because $F_{s}\left(\rho\right)=\left|\braket{\psi|\phi}\right|^{2}=\sum_{i=1}^{s}p_{i}F_{s}\left(\ket{\psi_{i}}\right)$,
which is exactly the optimality condition.

So far we proved the existence of an optimal decomposition $\left\{ p_{i},\ket{\psi_{i}}\right\} $
with the property $\sqrt{F_{s}\left(\ket{\psi_{i}}\right)}=\braket{\psi_{i}|\phi_{i}}$
starting from the existence of the closest separable state $\sigma=\sum_{j=1}^{s}q_{j}\ket{\phi_{j}}\bra{\phi_{j}}$.
Now we will prove the inverse direction. Given an optimal decomposition
$\left\{ p_{i},\ket{\psi_{i}}\right\} _{i=1}^{s}$ we will find a
closest separable state. We again define the purifications of $\rho$
and $\sigma$ as \begin{eqnarray}
\ket{\psi} & = & \sum_{i=1}^{s}\sqrt{p_{i}}\ket{\psi_{i}}\otimes\ket{i},\\
\ket{\phi} & = & \sum_{j=1}^{s}\sqrt{q_{j}}\ket{\phi_{j}}\otimes\ket{j},\end{eqnarray}
 where we define the states $\ket{\phi_{j}}$ to be separable and
to have maximal overlap with $\ket{\psi_{j}}$, i.e. $\braket{\psi_{j}|\phi_{j}}=\sqrt{F_{s}\left(\ket{\psi_{j}}\right)}$.
The real numbers $q_{j}$ are defined as follows: $q_{j}=\frac{p_{j}F_{s}\left(\ket{\psi_{j}}\right)}{\sum_{k}p_{k}F_{s}\left(\ket{\psi_{k}}\right)}$.
Now we note that $\left|\braket{\psi|\phi}\right|^{2}=F_{s}\left(\rho\right)$
because the decomposition $\left\{ p_{i},\ket{\psi_{i}}\right\} $
was defined to be optimal. Thus we see that there exists no purification
$\ket{\phi'}$ such that $\left|\braket{\psi|\phi'}\right|>\left|\braket{\psi|\phi}\right|$.
Together with Uhlmann's theorem this implies that $F\left(\rho,\sigma\right)=F_{s}\left(\rho\right)$.

\end{proof}

\subsection{Caratheodory bound}

Now we are in position to show that the number of elements in an optimal
decomposition (w.r.t. the geometric measure of entanglement) is bounded
from above by the Caratheodory bound. 
\begin{cor}
\label{cor:optimal decomposition}For any state $\rho$ acting on
a Hilbert space of dimension $d$ always exists an optimal (w.r.t.
the geometric measure of entanglement) decomposition $\left\{ p_{i},\ket{\psi_{i}}\right\} _{i=1}^{s}$
such that $s\leq d^{2}$.\end{cor}
\begin{proof}
Let $\sigma$ be a closest separable state. From Caratheodory's theorem
\cite{Horodecki1997,Vedral1998} follows that $\sigma$ can be written
as a convex combination of $s\leq d^{2}$ pure separable states. According
to Proposition \ref{pro:state-decomposition} the state $\sigma$
can be used to find an optimal decomposition with $s$ elements.
\end{proof}

\section{\label{sec:Structure}Structure of optimal decomposition w.r.t. geometric
measure of entanglement}

In this section we will show that the optimal decomposition of $\rho$
with respect to the geometric measure of entanglement has a certain
symmetric structure.

\subsection{\label{sub:n-partite-states}$n$-partite states}

First we derive the structure of an optimal decomposition $\left\{ p_{i},\ket{\psi_{i}}\right\} $
for a general $n$-partite state. 
\begin{prop}
\label{pro:structure} Every optimal decomposition $\left\{ p_{i},\ket{\psi_{i}}\right\} _{i=1}^{s}$
must have the following structure: \begin{equation}
\sqrt{F_{s}\left(\ket{\psi_{k}}\right)}\braket{\psi_{i}|\phi_{k}}=\sqrt{F_{s}\left(\ket{\psi_{i}}\right)}\braket{\phi_{i}|\psi_{k}}\label{eq:structure}\end{equation}
 for all $1\leq i,k\leq s$. Here the states $\ket{\phi_{i}}$ are
separable and have the property $\braket{\phi_{i}|\psi_{i}}=\sqrt{F_{s}\left(\ket{\psi_{i}}\right)}$.
\end{prop}
Eq. (\ref{eq:structure}) represent a nonlinear system of equations.
Finding all solutions of it is equivalent to computing the optimal
decomposition of $\rho$. For pure states our result reduces to the
nonlinear eigenproblem given in equations (5a) and (5b) in \cite{Wei2003}.

\begin{proof} Let the states $\ket{i}$ denote an orthonormal basis
on the ancillary Hilbert space $\mathcal{H}_{a}$. Let $\ket{\psi}=\sum_{i}\sqrt{p_{i}}\ket{\psi_{i}}\ket{i}$
and $\ket{\phi}=\sum_{j}\sqrt{q_{j}}\ket{\phi_{j}}\ket{j}$ be purifications
of $\rho$ and $\sigma$, respectively, such that $\left\{ p_{i},\ket{\psi_{i}}\right\} $
is an optimal decomposition of $\rho$, $\braket{\psi_{i}|\phi_{i}}=\sqrt{F_{s}\left(\ket{\psi_{i}}\right)}$
and $q_{i}=\frac{p_{i}F_{s}\left(\ket{\psi_{i}}\right)}{\sum_{k}p_{k}F_{s}\left(\ket{\psi_{k}}\right)}$.
This implies that \begin{equation}
F_{s}\left(\rho\right)=\left|\braket{\psi|\phi}\right|^{2}=\sum_{i}\left|\bra{\psi}\left(\ket{\phi_{i}}\otimes\ket{i}\right)\right|^{2}.\label{eq:Fs-1}\end{equation}
 Optimality implies that $\left|\braket{\psi|\phi}\right|^{2}$ is
stationary under unitaries acting on the ancillary Hilbert space $\mathcal{H}_{a}$
(for stationarity under unitaries acting on the original space see
subsection \ref{stationaryoriginal}), that is \begin{equation}
\frac{d}{dt}\left|\braket{\psi|e^{itH_{a}}|\phi}\right|_{t=0}^{2}=0\end{equation}
 for any Hermitian $H_{a}=H_{a}^{\dagger}$ acting on $\mathcal{H}_{a}$
and the derivative is taken at $t=0$. Using \eqref{eq:Fs-1} we can
write \begin{eqnarray}
\left|\braket{\psi|e^{itH_{a}}|\phi}\right|^{2} & = & \sum_{k}\left|\bra{\psi}\left(\ket{\phi_{k}}e^{itH_{a}}\ket{k}\right)\right|^{2}.\end{eqnarray}
 The derivative at $t=0$ becomes: \begin{equation}
\frac{d}{dt}\left|\braket{\psi|e^{itH_{a}}|\phi}\right|_{t=0}^{2}=\mathrm{Tr}_{a}\left[H_{a}\cdot\mathrm{Tr}_{\bar{a}}\left[\sum_{k}\left(A_{k}+A_{k}^{\dagger}\right)\right]\right]\label{eq:G}\end{equation}
 with $A_{k}=i\left(\ket{\phi_{k}}\bra{\phi_{k}}\otimes\ket{k}\bra{k}\right)\ket{\psi}\bra{\psi}$
and $\mathrm{Tr}_{\bar{a}}$ means partial trace over all parts except
for the ancillary space $\mathcal{H}_{a}$. Using $\left(\bra{\phi_{k}}\bra{k}\right)\ket{\psi}=\sqrt{p_{k}}\sqrt{F_{s}\left(\ket{\psi_{k}}\right)}$
we can write $A_{k}$ as \begin{equation}
A_{k}=i\sqrt{p_{k}F_{s}\left(\ket{\psi_{k}}\right)}\ket{\phi_{k}}\ket{k}\bra{\psi}.\end{equation}
 Expression \eqref{eq:G} has to be zero for all Hermitian $H_{a}$
which can only be true if $\mathrm{Tr}_{\bar{a}}\left[\sum_{k}\left(A_{k}+A_{k}^{\dagger}\right)\right]=0$
which is equivalent to \begin{eqnarray}
\sum_{k}\mathrm{Tr}_{\bar{a}}\left[\sqrt{p_{k}F_{s}\left(\ket{\psi_{k}}\right)}\ket{\phi_{k}}\ket{k}\bra{\psi}\right]\\
=\sum_{k}\mathrm{Tr}_{\bar{a}}\left[\sqrt{p_{k}F_{s}\left(\ket{\psi_{k}}\right)}\ket{\psi}\bra{k}\bra{\phi_{k}}\right].\nonumber \end{eqnarray}
 With $\ket{\psi}=\sum_{i}\sqrt{p_{i}}\ket{\psi_{i}}\ket{i}$ we get
\begin{eqnarray}
\sum_{i,k}\sqrt{p_{k}p_{i}F_{s}\left(\ket{\psi_{k}}\right)}\braket{\psi_{i}|\phi_{k}}\ket{k}\bra{i}\\
=\sum_{i,k}\sqrt{p_{i}p_{k}F_{s}\left(\ket{\psi_{k}}\right)}\braket{\phi_{k}|\psi_{i}}\ket{i}\bra{k}.\nonumber \end{eqnarray}
 Using orthogonality of $\left\{ \ket{i}\right\} $ completes the
proof. \end{proof}

\subsection{Bipartite states}

Let us illustrate the structure of an optimal decomposition with the
example of bipartite states. We consider the expression \eqref{eq:structure}
for a bipartite mixed state $\rho$ with optimal decomposition $\left\{ p_{i},\ket{\psi_{i}}\right\} $.
In this case it is possible to write the Schmidt decomposition of
the pure states $\ket{\psi_{i}}$ as follows:\begin{equation}
\ket{\psi_{i}}=\sum_{j}\lambda_{i,j}\ket{j_{i}^{\left(1\right)}}\ket{j_{i}^{\left(2\right)}}\label{eq:psii}\end{equation}
 with $\sum_{j}\lambda_{i,j}^{2}=1$, and the Schmidt coefficients
are in decreasing order, i.e. $\lambda_{i,1}\geq\lambda_{i,2}\geq...>0$.
The separable states $\ket{\phi_{i}}$ that have the highest overlap
with $\ket{\psi_{i}}$ are given by \[
\ket{\phi_{i}}=\ket{1_{i}^{\left(1\right)}}\ket{1_{i}^{\left(2\right)}},\]
 and $\sqrt{F_{s}\left(\ket{\psi_{i}}\right)}=\lambda_{i,1}$. With
this in mind expression \eqref{eq:structure} reduces to \begin{equation}
\lambda_{k,1}\braket{\psi_{i}|1_{k}^{\left(1\right)}}\ket{1_{k}^{\left(2\right)}}=\lambda_{i,1}\bra{1_{i}^{\left(1\right)}}\braket{1_{i}^{\left(2\right)}|\psi_{k}}\label{eq:structure-1}\end{equation}
 for all $i$, $k$.

\subsection{Qubit-qudit states}

Let now the first system be a qubit, that is $d_{1}=2$. In this case
we can set $\lambda_{k,1}=\cos\alpha_{k}$ and $\lambda_{k,2}=\sin\alpha_{k}$,
with $\cos\alpha_{k}\geq\sin\alpha_{k}$. With $\ket{\psi_{k}}=\cos\alpha_{k}\ket{11}+\sin\alpha_{k}\ket{22}$
we get from eq. (\ref{eq:structure-1}) \begin{eqnarray}
\cos\alpha_{k}\sin\alpha_{i}\left(\braket{2_{i}^{\left(1\right)}|1_{k}^{\left(1\right)}}\braket{2_{i}^{\left(2\right)}|1_{k}^{\left(2\right)}}\right)\\
=\cos\alpha_{i}\sin\alpha_{k}\left(\braket{1_{i}^{\left(1\right)}|2_{k}^{\left(1\right)}}\braket{1_{i}^{\left(2\right)}|2_{k}^{\left(2\right)}}\right).\nonumber \end{eqnarray}
 Noting that $\left|\braket{2_{i}^{\left(1\right)}|1_{k}^{\left(1\right)}}\right|=\left|\braket{1_{i}^{\left(1\right)}|2_{k}^{\left(1\right)}}\right|$
it follows that \begin{equation}
\frac{\tan\alpha_{i}}{\tan\alpha_{k}}=\left|\frac{\braket{1_{i}^{\left(2\right)}|2_{k}^{\left(2\right)}}}{\braket{2_{i}^{\left(2\right)}|1_{k}^{\left(2\right)}}}\right|.\label{eq:2}\end{equation}
 It is interesting to mention that in the case $d_{2}=2$ we can simplify
\eqref{eq:2} to $\tan\alpha_{i}=\tan\alpha_{k}$. This means that
in the optimal decomposition $\left\{ p_{i},\ket{\psi_{i}}\right\} $
of a two-qubit state all states $\ket{\psi_{i}}$ have the same Schmidt
coefficients, a result already known from \cite{Wootters1998}.

\subsection{Nonoptimal stationary decompositions}

Note that expression \eqref{eq:structure} is necessary, but not sufficient
for a decomposition to be optimal. To prove this we will give two
non-optimal decompositions that satisfy \eqref{eq:structure}.

\subsubsection{Bell diagonal states}

Consider the state \begin{equation}
\rho=\frac{1}{2}\ket{\psi^{+}}\bra{\psi^{+}}+\frac{1}{2}\ket{\phi^{+}}\bra{\phi^{+}},\label{eq:Bell}\end{equation}
 with $\ket{\psi^{+}}=\frac{1}{\sqrt{2}}\left(\ket{01}+\ket{10}\right)$
and $\ket{\phi^{+}}=\frac{1}{\sqrt{2}}\left(\ket{00}+\ket{11}\right)$.
It is well known that the state \eqref{eq:Bell} is separable, and
thus the decomposition into Bell states cannot be optimal. On the
other hand, it is easy to see that this decomposition satisfies \eqref{eq:structure}.

\subsubsection{Separable states}

Now we will give a more complicated example. We call a decomposition
$\left\{ p_{i},\ket{\psi_{i}}\right\} _{i=1}^{s}$ $s$-optimal if
for a given number of terms $s$ there is no decomposition $\left\{ q_{i},\ket{\phi_{i}}\right\} _{i=1}^{s}$
such that $\sum_{i=1}^{s}q_{i}E_{G}\left(\ket{\phi_{i}}\right)<\sum_{i=1}^{s}p_{i}E_{G}\ket{\psi_{i}}$.
It is known \cite{Horodecki2009} that there exist separable states
$\rho$ of dimension $d$ with the property that any $d$-optimal
decomposition is not separable and thus not optimal. Let $\left\{ p_{i},\ket{\psi_{i}}\right\} _{i=1}^{d}$
be a $d$-optimal decomposition of such a state $\rho$.

We write a purification of $\rho$ as $\ket{\psi}=\sum_{i=1}^{d}\sqrt{p_{i}}\ket{\psi_{i}}\ket{i}$.
Further we define separable states $\ket{\phi_{i}}$ such that $\braket{\psi_{i}|\phi_{i}}=\sqrt{F_{s}\left(\ket{\psi_{i}}\right)}$,
$q_{i}=\frac{p_{i}F_{s}\left(\ket{\psi_{i}}\right)}{\sum_{k}p_{k}F_{s}\left(\ket{\psi_{k}}\right)}$
and $\ket{\phi}=\sum_{j=1}^{d}\sqrt{q_{j}}\ket{\phi_{j}}\ket{j}$.
Then it holds that: \begin{equation}
\left|\braket{\psi|\phi}\right|^{2}=\sum_{i=1}^{d}p_{i}F_{s}\left(\ket{\psi_{i}}\right)^{2}.\end{equation}
 From $d$-optimality of $\left|\braket{\psi|\phi}\right|^{2}$ follows
that for all Hermitian matrices acting on a $d$-dimensional Hilbert
space $\mathcal{H}_{a}$ \begin{equation}
\frac{d}{dt}\left|\braket{\psi|e^{itH_{a}}|\phi}\right|_{t=0}^{2}=0\label{eq:nonoptimal}\end{equation}
 holds. We will now show that $\frac{d}{dt}\left|\braket{\psi|e^{itH_{a}}|\phi}\right|_{t=0}^{2}=0$
also holds for dim$\left(\mathcal{H}_{a}\right)\geq d$. This means
that adding more dimensions to the ancillary Hilbert space will not
help. Doing the same calculation as in the proof of Proposition \ref{pro:structure}
we get: \begin{equation}
\frac{d}{dt}\left|\braket{\psi|e^{itH_{a}}|\phi}\right|_{t=0}^{2}=\mathrm{Tr}_{a}\left[H_{a}\cdot\mathrm{Tr}_{\bar{a}}\left[\sum_{k=1}^{d\left(\mathcal{H}_{a}\right)}\left(A_{k}+A_{k}^{\dagger}\right)\right]\right]\end{equation}
 with $A_{k}=i\sqrt{p_{k}F_{s}\left(\ket{\psi_{k}}\right)}\ket{\phi_{k}}\ket{k}\bra{\psi}$.
Note that $A_{k}$ is nonzero only for $k\leq d$, because $p_{k}=0$
otherwise. Thus we can restrict ourselves to $k\leq d$ in the calculation,
which is equivalent to setting dim$\left(\mathcal{H}_{a}\right)=d$.
Then \eqref{eq:nonoptimal} implies $\mathrm{Tr}_{\bar{a}}\left[\sum_{k=1}^{d\left(\mathcal{H}_{a}\right)}\left(A_{k}+A_{k}^{\dagger}\right)\right]=0$
and it follows that \eqref{eq:nonoptimal} holds for arbitrary $d\left(\mathcal{H}_{a}\right)\geq d$.

\subsection{Stationarity on the original subspace}

\label{stationaryoriginal} In Proposition \ref{pro:structure} we
used the argument that in the optimal case $\left|\braket{\psi|\phi}\right|^{2}$
has to be stationary under unitaries acting on the ancillary Hilbert
space $\mathcal{H}_{a}$. In \eqref{eq:Fs-1} we could rewrite this
expression as \[
F_{s}\left(\rho\right)=\left|\braket{\psi|\phi}\right|^{2}=\sum_{i}\left|\braket{\psi|\phi_{i}}\ket{i}\right|^{2},\]
 where all $\ket{\phi_{i}}$ are separable. We can also demand $\sum_{i}\left|\braket{\psi|\phi_{i}}\ket{i}\right|^{2}$
to be stationary under (separable) unitaries acting on the original
Hilbert space of the states $\ket{\phi_{i}}$. From this procedure
we will gain stationary equations describing the states $\ket{\phi_{i}}$.
However, we already know that in the optimal case we can choose $\ket{\phi_{i}}$
to be the closest separable state to $\ket{\psi_{i}}$, that is $\braket{\psi_{i}|\phi_{i}}=\sqrt{F_{s}\left(\ket{\psi_{i}}\right)}$,
such that this method does not give new results.

\section{\label{sec:Concluding-remarks}Concluding remarks}

We have shown in this paper that the geometric measure of entanglement
belongs to two classes of entanglement measures. Namely it is a convex
roof measure and also a distance measure of entanglement. As an application
we gave a closed formula for the Bures measure of entanglement for
two qubits. We also note that the revised geometric measure of entanglement
defined in \cite{Cao2007} is equal to the original geometric measure
of entanglement.

We furthermore proved that the problems of finding a closest separable
state and finding an optimal decomposition are equivalent. We used
this insight to bound the number of elements in an optimal decomposition
(with respect to the geometric measure of entanglement). It turns
out that the bound is exactly given by the Caratheodory bound.

Finally, we obtained stationary equations which ensure optimality
of a decomposition. For the case of two qubits these equations lead
to the known fact that each constituting state of an optimal decomposition
has equal concurrence. Our equations hold for any dimension. However,
they are only necessary, not sufficient for a decomposition to be
optimal. Given an arbitrary decomposition, they provide a simple test
whether the decomposition may be optimal. 
\begin{acknowledgments}
We acknowledge discussion with M. Plenio. A. S. also thanks C. Gogolin,
H. Hinrichsen, and P. Janotta. This work was partially supported by
DFG (Deutsche Forschungsgemeinschaft). 
\end{acknowledgments}
\appendix

\section{\label{sec:Convex-sets}Geometric measure of a convex set}

In Theorem \ref{thm:main} we stated that if $S$ is the set of separable
states it holds:\begin{equation}
F_{s}\left(\rho\right)=\max\sum_{i}p_{i}F_{s}\left(\ket{\psi_{i}}\right),\end{equation}
 where $F_{s}$ is the maximal fidelity between $\rho$ and the set
of separable states: $F_{s}\left(\rho\right)=\max_{\sigma\in S}F\left(\rho,\sigma\right)$
and the maximisation is done over all pure state decompositions of
$\rho$. In the following we will generalise this result to arbitrary
convex sets.

Let $X$ be a set of states $\left\{ \sigma_{k}\right\} $ and $C$
be a set containing all convex combinations of the elements of $X$,
these are states $\sigma$ such that holds: \begin{equation}
\sigma=\sum_{k}q_{k}\sigma_{k}\end{equation}
 with $q_{k}\geq0$, $\sum_{k}q_{k}=1$. We define the quantities
$F_{X}\left(\rho\right)$ and $F_{C}\left(\rho\right)$ to be the
maximal fidelity between $\rho$ and an element of $X$ and $C$ respectively:
\begin{eqnarray}
F_{X}\left(\rho\right) & = & \max_{\sigma\in X}F\left(\rho,\sigma\right),\\
F_{C}\left(\rho\right) & = & \max_{\sigma\in C}F\left(\rho,\sigma\right).\end{eqnarray}

\begin{thm} For an arbitrary quantum state $\rho$ and a convex set
of states $C$ holds \begin{eqnarray}
F_{C}\left(\rho\right) & = & \max_{\rho=\sum_{k}p_{k}\rho_{k}}\sum_{i}p_{i}F_{X}\left(\rho_{i}\right),\end{eqnarray}
 where the maximisation is done over all decompositions of $\rho=\sum_{i}p_{i}\rho_{i}$,
$p_{i}\geq0$.\end{thm} \begin{proof} The proof is a modification
of the proof of Theorem \ref{thm:main}. According to Uhlmann's theorem
\cite[page 411]{Nielsen2000} holds: \begin{equation}
F\left(\rho,\sigma\right)=\max_{\ket{\phi}}\left|\braket{\psi|\phi}\right|^{2},\end{equation}
 $\ket{\psi}$ is a purification of $\rho$ and the maximisation is
done over all purifications of $\sigma$ denoted by $\ket{\phi}$.

In order to find $F_{C}\left(\rho\right)$ we have to maximise $\left|\braket{\psi|\phi}\right|^{2}$
over purifications $\ket{\phi}$ of all states of the form $\sigma=\sum_{k}q_{k}\sigma_{k}$,
$\sigma_{k}\in X$. Using similar arguments as in the proof of the
Theorem \ref{thm:main} we see that the purifications can always be
written as \begin{eqnarray}
\ket{\psi} & = & \sum_{i}\sqrt{p_{i}}\left(\sum_{j}\sqrt{p_{i,j}}\ket{\psi_{i,j}}\otimes\ket{i,j}\right),\\
\ket{\phi} & = & \sum_{k}\sqrt{q_{k}}\left(\sum_{l}\sqrt{q_{k,l}}\ket{\phi_{k,l}}\otimes\ket{k,l}\right),\end{eqnarray}
 with $\braket{i,j|k,l}=\delta_{ik}\delta_{jl}$. In the maximisation
of $\left|\braket{\psi|\phi}\right|^{2}$ we are free to choose the
states $\ket{\phi_{k,l}}$ under the restriction that $\sum_{l}\sqrt{q_{k,l}}\ket{\phi_{k,l}}\otimes\ket{k,l}$
purifies $\sigma_{k}\in X$, the probabilities $q_{k}>0$ are restricted
only by $\sum_{k}q_{k}=1$. We are also free to choose $\left\{ \ket{\psi_{i,j}}\right\} $,
$\left\{ p_{i}\right\} $ and $\left\{ p_{i,j}\right\} $ under the
restriction $\rho=\sum_{i,j}p_{i}p_{ij}\ket{\psi_{i,j}}\bra{\psi_{i,j}}$.
With this in mind we get: \begin{equation}
\left|\braket{\psi|\phi}\right|=\left|\sum_{i,k}\sqrt{p_{i}q_{k}}a_{i,k}\right|,\end{equation}
 with $a_{i,k}$ being the product of the purifications of $\rho_{i}$
and $\sigma_{k}$: \[
a_{i,k}=\left(\sum_{j}\sqrt{p_{i,j}}\bra{\psi_{i,j}}\otimes\bra{i,j}\right)\left(\sum_{l}\sqrt{q_{k,l}}\ket{\phi_{k,l}}\otimes\ket{k,l}\right).\]
 Now we optimise over $\left\{ q_{k,l},\ket{\phi_{k,l}}\right\} $
with the result \begin{equation}
a_{i,k}=\sqrt{F_{X}\left(\rho_{i}\right)}\delta_{ik}\end{equation}
 and thus \begin{equation}
\max_{\left\{ q_{k,l},\ket{\phi_{k,l}}\right\} }\left|\braket{\psi|\phi}\right|=\sum_{i}\sqrt{q_{i}p_{i}}\sqrt{F_{X}\left(\rho_{i}\right)}.\end{equation}
 Now we do the optimisation over $q_{i}$. Using Lagrange multipliers
we get \begin{equation}
\sqrt{q_{i}}=\frac{\sqrt{p_{i}}\sqrt{F_{X}\left(\rho_{i}\right)}}{\sqrt{\sum_{k}p_{k}F_{X}\left(\rho_{k}\right)}},\end{equation}
 with the result \begin{equation}
\max_{\left\{ q_{j},q_{k,l},\ket{\phi_{k,l}}\right\} }\left|\braket{\psi|\phi}\right|^{2}=\sum_{i}p_{i}F_{X}\left(\rho_{i}\right).\end{equation}
 In the last step we do the maximisation over all decompositions $\left\{ p_{i},\rho_{i}\right\} $
of the given state $\rho$ which leads to the final result \begin{equation}
F_{C}\left(\rho\right)=\max\left|\braket{\psi|\phi}\right|^{2}=\max\sum_{i}p_{i}F_{X}\left(\rho_{i}\right).\end{equation}

\end{proof} \bibliographystyle{h-physrev} \bibliographystyle{h-physrev}
\bibliographystyle{apsrev4-1}
\bibliography{literature}

\end{document}